\documentstyle[aps,prl,twocolumn,graphicx,floats,epsfig,times,here]{revtex}

\begin{document}

\draft 

\onecolumn 

\noindent

\title{
Localized Charged-Neutral Fluctuations in
  158$~${\it A}~GeV Pb+Pb Collisions}

\author{ M.M.~Aggarwal,$^{1}$ A.~Agnihotri,$^{2}$ Z.~Ahammed,$^{3}$
  A.L.S.~Angelis,$^{4}$ V.~Antonenko,$^{5}$ V.~Arefiev,$^{6}$
  V.~Astakhov,$^{6}$ V.~Avdeitchikov,$^{6}$ T.C.~Awes,$^{7}$
  P.V.K.S.~Baba,$^{8}$
% ...
}

\author{(WA98 Collaboration)}

\address{$^{1}$~University of Panjab, Chandigarh 160014, India}
\address{$^{2}$~University of Rajasthan, Jaipur 302004, Rajasthan,
  India} \address{$^{3}$~Variable Energy Cyclotron Centre, Calcutta
  700 064, India} \address{$^{4}$~University of Geneva, CH-1211 Geneva
  4, Switzerland} \address{$^{5}$~RRC ``Kurchatov Institute'', 
  RU-123182 Moscow,
  Russia} \address{$^{6}$~Joint Institute for Nuclear Research,
  RU-141980 Dubna, Russia} \address{$^{7}$~Oak Ridge National
  Laboratory, Oak Ridge, Tennessee 37831-6372, USA}
\address{$^{8}$~University of Jammu, Jammu 180001, India}
% ...

\date{Draft 5.0, \today} \maketitle
\begin{abstract}
  First results on the measurement of localized fluctuations in the
  multiplicity of charged particles and photons produced in central
  158$\cdot A$~GeV/c Pb+Pb collisions are presented.  The charged
  versus neutral correlations in common phase space regions of varying
  azimuthal size are analyzed by two different methods. The measured
  results are compared to those from simulations and to those from
  different types of mixed events.  The comparison indicates the
  presence of non-statistical fluctuations in both charged particle
  and photon multiplicities in limited azimuthal regions.  However, no
  correlated charge-neutral fluctuations are observed.

\end{abstract}
\twocolumn


  The formation of hot and dense matter in high energy heavy-ion
  collisions offers the possibility to create a new phase where matter
  is deconfined and chiral symmetry is restored.  Indications for the
  formation of such a Quark Gluon Plasma (QGP) phase are provided by
  several results from experiments at the CERN-SPS \cite{qm99}.
  Event-by-event fluctuations in the particle multiplicities and their
  ratios have recently been predicted to provide information about the
  nature of the QCD phase transition \cite{stephanov,jeon}.
  Fluctuations may also be caused by Bose-Einstein correlations,
  resonance decays, or more exotic phenomena such as pion lasers
  \cite{pratt}.  Enhanced fluctuations in neutral to charged pions
  have been predicted as a signature of the formation of disoriented
  chiral condensates (DCC) \cite{anselm,bj,blaizot,raj}, which might
  be one of the most interesting predicted consequences of chiral
  symmetry restoration.

  Theoretical predictions suggest that isospin fluctuations, caused by
  formation of a DCC, would produce clusters of coherent pions in
  localized phase space regions or domains. The probability
  distribution of the neutral pion fraction in such a domain would
  follow the relation $P(f) = 1 / 2 \sqrt{f}$, where $f =
  N_{\pi^0}/N_{\pi}$.  Thus DCC formation in a given domain would be
  associated with large event-by-event fluctuations in the ratio of
  neutral to charged pions in that domain. Experimentally, such
  fluctuations can be deduced from the measurement of fluctuations in
  the number of photons to charged particles in limited $\eta$--$\phi$
  regions.  The anti-Centauro events, reported by the JACEE
  collaboration \cite{jacee}, with large charged-neutral fluctuations
  are possible candidates for DCC events.  The studies carried out so
  far in $p-\bar{p}$ \cite{minimax} and heavy ion \cite{WA98-3,NA49}
  reactions have searched for fluctuations which extend over a large
  region of phase space.  These measurements have provided upper
  limits on the presence of DCC-like fluctuations.

  In this letter we present first results on the search for
 non-statistical event-by-event fluctuations in the relative number of
 charged particles and photons in localized phase space regions for
 central 158$\cdot A$~GeV/c Pb+Pb collisions.  The data presented here
 were taken with the 158$\cdot A$~GeV Pb beam of the CERN SPS on a Pb
 target of 213~$\mu$m thickness during a period of WA98 operation
 without magnetic field.  The analysis makes use of a subset of
 detectors of the WA98 experiment which are used to measure the
 multiplicities of charged particles and photons. Charged particle
 hits ($N_{\mathrm ch}$) were counted using a circular Silicon Pad
 Multiplicity Detector (SPMD) \cite{WA98-3} located 32.8 cm downstream
 from the target. It provided uniform pseudorapidity coverage in the
 region $2.35< \eta < 3.75$. The detector was 99$\%$ efficient for
 charged particle detection.  The photon multiplicity was measured
 using a preshower Photon Multiplicity Detector (PMD) \cite{wa98nim}
 placed 21.5 meters downstream of the target and covering the
 pseudorapidity range $2.9< \eta < 4.2$.  It consisted of an array of
 53,200 plastic scintillator pads placed behind 3$X_0$ thick lead
 converter plates. Clusters of hit pads having total energy deposit
 above a hadron rejection threshold are identified as
 photon${-{\mathrm like}}$. The photon counting efficiency and the
 purity of the ${\gamma-{\mathrm like}}$ sample have been found to be
 68$\%$ and 65$\%$, respectively, for central events \cite{WA98-9}.
 For this analysis the pseudorapidity region of common coverage of the
 SPMD and PMD was selected ($2.9< \eta < 3.75$).  The acceptance in
 terms of transverse momentum ($p_T$) extends down to 30 MeV/c,
 although no explicit $p_T$-selection is applied.  Events with pile-up
 or downstream interactions were rejected in the off-line analysis.
 Strict data selection and cleanup cuts have been applied as described
 in Ref.~\cite{WA98-3,WA98-9}. After cuts, a total of 85K central
 events, corresponding to the top 5$\%$ of the minimum bias cross
 section as determined from the measured total transverse energy, have
 been analyzed.

 The measured results are interpreted by comparison with simulated
    events and with several types of mixed events.  Simulated events
    were generated using the VENUS 4.12 \cite{venus} event generator
    with default parameters. The output was processed through a WA98
    detector simulation package in the GEANT 3.21 \cite{geant}
    framework.  The centrality selection for the simulated data has
    been made in an identical manner to the experimental data by
    selection on the simulated total transverse energy in the WA98
    acceptance.  The simulated VENUS+GEANT events (referred as V+G)
    were then processed with the same analysis codes as used for the
    analysis of the experimental data.

  The effect of non-statistical DCC-like charged-neutral fluctuations
has been studied within the framework of a simple model in which the
output of the VENUS event generator has been modified.  It is expected
that DCC domains will occur in small regions and will mostly modify
the production of low momentum pions.  The influence of a DCC on the
charge-neutral pion ratio may then be limited to localized
$\eta$--$\phi$ regions due to the motion of the DCC domain within the
overall collective motion.  To implement the DCC effect, the charges
of the pions within a localized $\eta$--$\phi$ region predicted by
VENUS are interchanged pairwise ($\pi^{+}\pi^{-} \leftrightarrow
\pi^{0}\pi^{0}$) according to the $1 / 2 \sqrt{f}$ probability
distribution.  The DCC-like fluctuations were generated over
$\eta=3-4$ for varying intervals in $\Delta\phi$.  Since the
probability to produce events with DCC domains is unknown, ensembles
of events, here referred to as a ``nDCC events'', were produced which
consisted of a mixture of normal events with varying fractions of pure
DCC-like events.  The nDCC events were then tracked through GEANT.

In the search for evidence of non-statistical charged-neutral
fluctuations, two different analysis techniques have been applied.
The first method employed in the present analysis is the technique of
discrete wavelet transformations (DWT).  DWT methods are now widely
used in many applications, such as data compression and image
processing, and have been shown to provide a powerful means to search
for localized domains of DCC \cite{huang,dccstr}.  While there are
several families of wavelet bases distinguished by the number of
coefficients and the level of iteration, we have used the frequently
employed $D-$4 wavelet basis \cite{numeric}, which are orthogonal,
continuously differentiable and localized in space. The analysis has
been performed with the sample function chosen to be the photon
fraction, given by $f^\prime(\phi) = {N_{\gamma-{\mathrm
like}}(\phi)}/ {(N_{\gamma-{\mathrm like}}(\phi)+N_{\mathrm
ch}(\phi))}$ as a function of the azimuthal angle $\phi$, with highest
resolution scale $j_{max}=5$. The input to the DWT analysis is the
spectrum of the sample function at the smallest bin size corresponding
to the highest resolution scale, $j_{max}$, where the number of bins
is $2^{j_{max}}$. The sample function is then analyzed at different
scales $j$ by being re-binned into $2^j$ bins.  The DWT analysis
yields a set of wavelets or father function coefficients (FFC) at each
scale from $j=1$,...,($j_{max}-1$).  The coefficients obtained at a
given scale, $j$, are derived from the distribution of the sample
function at one higher scale, $j+1$.  The FFCs quantify the deviation
of the bin-to-bin fluctuations in the sample function at that higher
scale relative to the average behavior. The presence of localized
non-statistical fluctuations will increase the rms deviation of the
distribution of FFCs and may result in non-Gaussian tails
\cite{huang,dccstr}.

\begin{figure}
\begin{center}
\vspace*{-0.9cm}
\epsfig{figure=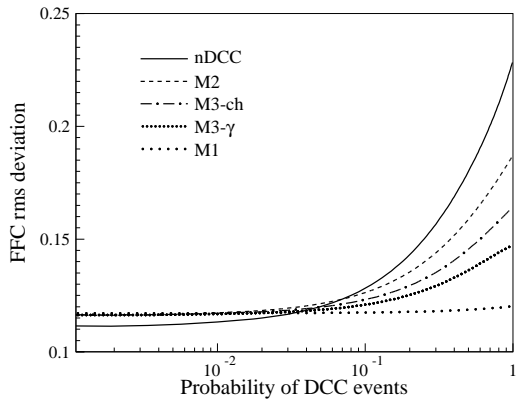,width=7.5cm}
\caption {\label{m1_m2_check}
The rms deviations of the FFC distributions at $j=1$ for
simulated nDCC events with extent $\Delta\phi_{DCC}=90^\circ$ 
and for various mixed events constructed from those events, as
a function of the fraction of DCC-like events present 
in the nDCC sample.
}
\end{center}
\end{figure}

The DWT technique as used in this analysis is demonstrated in
Fig.~\ref{m1_m2_check} where it has been applied to the simulated nDCC
events. The rms deviation of the FFC distribution is shown as a
function of the fraction of DCC-like events in the nDCC sample. The
rms deviation is observed to increase strongly with increasing
DCC-like fraction.  Due to the inherent uncertainties in the
description of ``normal'' physics and detector response in the V+G
simulations, the observation of an experimental result with rms which
differs from the case with zero DCC fraction cannot be taken alone as
evidence of DCC observation.  For this reason four different types of
mixed events have been created from the real or simulated events in
order to search for non-statistical fluctuations by removing various
correlations in a controlled manner while preserving the
characteristics of the measured distributions as accurately as
possible.  The first type of mixed events (M1), are generated by
mixing hits in both the PMD and SPMD separately, with no two hits
taken from the same event. Hits within a detector in the mixed events
are not allowed to lie within the two track resolution of that
detector.  The second kind of mixed events (M2) are generated by
mixing the unaltered PMD hits of one event with the unaltered SPMD
hits of a different event. Intermediate between the M1 and M2 kinds of
mixed events is the case where the hits within the PMD are unaltered
while the SPMD hits are mixed (M3-$\gamma$), or the SPMD hits are
unaltered while the PMD hits are mixed (M3-$\mathrm ch$).  In each
type of mixed event the global (bin 1) $N_{\gamma-{\mathrm
like}}$--$N_{\mathrm ch}$ correlation is maintained as in the real
event.

   The rms deviations of the FFCs for the different kinds of mixed
events produced from the nDCC events are also shown in
Fig.~\ref{m1_m2_check}. In the case of vanishing DCC-like
fluctuations, the rms values of the various types of mixed events are
very close to each other.  The V+G rms values are lower than those of
the mixed events due to the presence of additional correlations
between $N_{\mathrm ch}$ and $N_{\gamma-{\mathrm like}}$ mostly as a
result of the charged particle contamination in the
$N_{\gamma-{\mathrm like}}$ sample.  The rms deviations for the M1
events are found to be almost independent of probability of DCC-like
events, while the rms deviations of the M2 events increase similarly,
but more weakly, than those of the nDCC events. The rms deviations for
the M3 sets of events are found to lie between M2 and M1. Thus, the
sequence of the mixed events relative to the simulated events (or
data) gives a model independent indication of the presence and source
of non-statistical fluctuations. The simple DCC model used here
results in an anti-correlation between $N_{\gamma-{\mathrm like}}$ and
$N_{\mathrm ch}$ due to the ``isospin-flip'' procedure used to
implement the DCC effect. It also results in non-statistical
fluctuations in both $N_{\gamma-{\mathrm like}}$ and $N_{\mathrm
ch}$. Thus the M2 events remove only the $N_{\gamma-{\mathrm
like}}$--$N_{\mathrm ch}$ anti-correlation while the M1 events are
seen to remove all non-statistical fluctuations and correlations. The
M3 mixed events give intermediate results because they contain only
the $N_{\gamma-{\mathrm like}}$ (M3-$\gamma$) or $N_{\mathrm ch}$
(M3-${\mathrm ch}$) non-statistical fluctuations.

\begin{figure}[t]
\begin{center}
\vspace*{-1.4cm}
\epsfig{figure=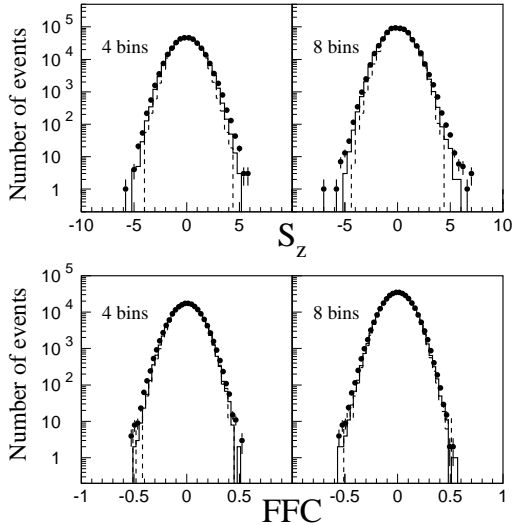,width=8.5cm}
\caption {\label{sz-ffc}
The $S_Z$ and FFC distributions for 4 and 8 divisions in $\phi$.
The experimental data, M1 and V+G events are
shown by solid circles, solid histograms and dashed histograms,
respectively.
The statistics for data and mixed events are same, whereas the
distribution for the V+G events is normalized to the number of data
events.
}
\end{center}
\end{figure}

   The FFC distributions extracted from the measured $f^\prime(\phi)$
ratio are shown in the bottom panel of Fig.~\ref{sz-ffc} for the
experimental data, for M1 events (from data), and for V+G events. The
results are shown for scales $j=$1 and 2, which carry information
about fluctuations at $90^\circ$ and $45^\circ$ in $\phi$.  The FFC
distributions of the experimental distributions are seen to be broader
than the V+G and M1 results.  This suggests the presence of
non-statistical fluctuations.

   A more conventional method similar to that described in
   Ref.~\cite{WA98-3} has also been used to search for non-statistical
   fluctuations.  The correlation between $N_{\gamma-{\mathrm like}}$
   and $N_{\mathrm ch}$ has been studied in varying $\phi$ intervals,
   by dividing the entire $\phi$-space into 2, 4, 8, and 16 bins.  The
   correlation plot of $N_{\mathrm ch}$ versus $N_{\gamma-{\mathrm
   like}}$ \cite{bedanga} is obtained for each $\phi$ segmentation,
   starting with the case of 1 bin which corresponds to the full
   azimuth.  A common correlation axis ($Z$) has been obtained for the
   full distribution by fitting the mean $N_{\gamma-{\mathrm like}}$
   and $N_{\mathrm ch}$ values with a second order polynomial.  The
   distance of separation ($D_{Z}$) between the data points and the
   correlation axis has been calculated with the convention that
   $D_{Z}$ is positive for points below the $Z$-axis (increasing
   $N_{\gamma-{\mathrm like}}$).  The distribution of $D_Z$ represents
   the fluctuations of $N_{\gamma-{\mathrm like}}$ relative to
   $N_{\mathrm ch}$ compared to the common correlation axis.  In order
   to compare fluctuations at different bin sizes having different
   multiplicities we use a scaled variable, $S_Z = D_Z/s(D_Z)$, where
   $s(D_Z)$ is the rms deviation of the $D_Z$ distributions for V+G
   events.  The presence of events with localized non-statistical
   fluctuations would be expected to result in a broader distribution
   of $S_Z$ compared to those for normal events. The $S_{Z}$
   distributions calculated at 4 and 8 bins in $\phi$ angle are shown
   in the top panel of Fig.~\ref{sz-ffc} for data, M1, and V+G
   events. The experimental distributions are broader than the
   simulation and M1 results, again indicating the presence of
   additional fluctuations.

\begin{figure}[t]
\begin{center}
\vspace*{-0.9cm}
\epsfig{figure=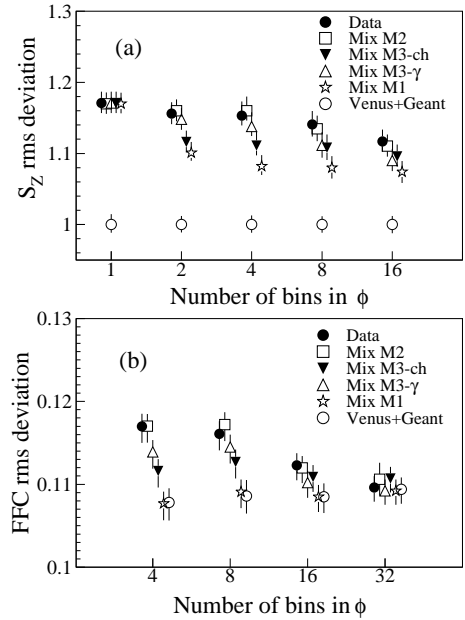,width=7cm}
\caption {\label{sz_ffc_rms}
The root mean square (rms) deviations of the $S_Z$  and FFC distributions
for various divisions in the azimuthal angle.
}
\end{center}
\end{figure}

   The rms deviations of the $S_Z$ and FFC distributions as a function
   of the number of bins in azimuth is shown for experimental data,
   mixed events, and V+G in Fig.~\ref{sz_ffc_rms}.  The statistical
   errors on the values are small and lie within the size of the
   symbols. The error bars include both statistical and systematic
   errors.  The systematic errors have been estimated by investigation
   of effects such as the uncertainties in the detection efficiencies,
   gain fluctuations, backgrounds, binning variations, and fitting
   procedures.  Since the mixed events are constructed to maintain the
   $N_{\gamma-{\mathrm like}}$--$N_{\mathrm ch}$ correlations for the
   full azimuth (bin~1), the rms deviations of data and mixed events
   for this bin are identical.  The difference of the $S_Z$ rms deviations
   between data and V+G for this bin is the same as reported in an
   earlier WA98 publication \cite{WA98-3}.  The comparison of 
   V+G and the M1 mixed events demonstrates the utility of the DWT
   method to normalize out the average behavior when the bin-to-bin
   fluctuation information is extracted.

As noted in the discussion of Fig.~\ref{m1_m2_check}, even in the
absence of DCCs there exist uninteresting correlations between
$N_{\gamma-{\mathrm like}}$ and $N_{\mathrm ch}$ which are removed by
the event mixing procedure and thereby result in a difference between
the real and mixed events. The various mixed event rms values of
Fig.~\ref{sz_ffc_rms} have therefore been rescaled by the percentage
difference between the rms deviations of the V+G distributions and
those of the corresponding V+G mixed events in order to better
illustrate effects in the data beyond those present in V+G.  Taking
the larger of the asymmetric systematic error at each point to be one
sigma, we find that at 2, 4, and 8 bins the values of the $S_Z$ rms
deviations of the data are 3.2$\sigma$, 3.4$\sigma$, and 3.1$\sigma$
larger than those of M1 events, respectively.  Similarly, the FFC rms
deviations at 4 and 8 bins for data are 4.3$\sigma$ and 3.3$\sigma$
larger than those of the M1 events.  At 16 and 32 bins, the result for
mixed events and data agree within the quoted errors.  The corrected
rms deviations of the M2 events agree with those of the experimental
data within error for all bins.  The M3 type mixed events are found to
be similar to each other within the quoted errors and lie between M1
and M2.

   The observation that the rms deviations of the $S_Z$ and FFC
   distributions for experimental data are larger than those of the M1
   events indicates the presence of localized non-statistical
   fluctuations. However, the comparision of the rms deviations for
   data with those of M2 events implies the absence of event-by-event
   correlated fluctuations in $N_{\gamma-{\mathrm like}}$ versus
   $N_{\mathrm ch}$.  The M3-type mixed events indicate the presence
   of localized independent fluctuations in $N_{\gamma-{\mathrm
   like}}$ and $N_{\mathrm ch}$ of similar magnitude.

   If the amount of DCC-like fluctuations in the experimental data
   were large, then the rms deviations shown in Fig. \ref{sz_ffc_rms}
   for data would have been larger compared to those of M2 events.
   Since this is not the case, we compare the measured results with
   those obtained from the simulation as shown in
   Fig. \ref{m1_m2_check} to extract upper limits on the probability
   of DCC-like fluctuations at the 90$\%$ confidence level. Within the
   context of this simple DCC model, upper limits on the presence of
   localized non-statistical DCC-like fluctuations of $10^{-2}$ for
   $\Delta\phi$ between 45--90$^\circ$ and $3\times 10^{-3}$ for
   $\Delta\phi$ between 90--135$^\circ$ are extracted.

    In summary, a detailed event-by-event analysis of the fluctuations
in the $\eta-\phi$ phase space distributions of charged particles and
photons has been performed for central Pb+Pb collisions at 158$\cdot
A$~GeV using two complementary analysis methods.  The first analysis
employed the discrete wavelet transformation technique to investigate
the relative magnitude of the $N_{\gamma-{\mathrm like}}$ versus
$N_{\mathrm ch}$ fluctuations in adjacent phase space regions.  The
second method studied the magnitude of the $N_{\gamma-{\mathrm like}}$
versus $N_{\mathrm ch}$ multiplicity fluctuations in decreasing phase
space regions.  The results were compared to pure VENUS+GEANT
simulations and to various types of mixed events to isolate the source
of non-statistical fluctuations. Both analysis methods indicated
non-statistical fluctuations beyond simulation and beyond pure mixed
events at the 3-4$\sigma$ level for $\phi$ intervals of greater than
45$^\circ$.  This is found to be due to uncorrelated non-statistical
fluctuations in $N_{\gamma-{\mathrm like}}$ and $N_{\mathrm ch}$. No
significant correlated fluctuations in $N_{\gamma-{\mathrm like}}$
versus $N_{\mathrm ch}$ were observed.  The results allow to set an
upper limit on the frequency of production of DCCs of limited domain
size, as demonstrated with a simple model of DCC-like fluctuations.

{
We wish to express our gratitude to the CERN accelerator division for the
excellent performance of the SPS accelerator complex. We acknowledge with
appreciation the effort of all engineers, technicians and support staff who
have participated in the construction of this experiment. 
This work was supported jointly by
the German BMBF and DFG,
the U.S. DOE,
the Swedish NFR and FRN,
the Dutch Stichting FOM,
the Polish KBN under contract 621/E-78/SPUB/CERN/P-03/DZ211,
the Grant Agency of the Czech Republic under contract No. 202/95/0217,
the Department of Atomic Energy,
the Department of Science and Technology,
the Council of Scientific and Industrial Research and
the University Grants
Commission of the Government of India,
the Indo-FRG Exchange Program,
the PPE division of CERN,
the Swiss National Fund,
the INTAS under Contract INTAS-97-0158,
ORISE,
Grant-in-Aid for Scientific Research
(Specially Promoted Research \& International Scientific Research)
of the Ministry of Education, Science and Culture,
the University of Tsukuba Special Research Projects, and
the JSPS Research Fellowships for Young Scientists.
ORNL is managed by UT-Battelle, LLC, for the U.S. Department of Energy
under contract DE-AC05-00OR22725.
The MIT group has been supported by the US Dept. of Energy under the
cooperative agreement DE-FC02-94ER40818.}

\end{document}